# Flagged observation analyses as a tool for scoping and communication in Integrated Ecosystem Assessments

Solvang, Hiroko Kato

Institute of Marine Research, P.O. Box 1870 Nordnes, N-5817 Bergen, Norway

hirokos@hi.no

Arneberg, Per

Institute of Marine Research, Fram Centre, P.O. Box 6606, 9296 Langnes, Norway

perab@hi.no

**Abstract**

Working groups for integrated ecosystem assessments are often challenged with understanding and assessing recent change in ecosystems. As a basis for this, the groups typically have at their disposal many time series and will often need to prioritize which ones to follow up for closer analyses and assessment. Here we provide a procedure termed Flagged observation analyses that can be applied to all the available time series to help identifying time series that should be prioritized. The statistical procedure first applies a structural time series model including a stochastic trend model to the data to estimate the long-term trend. The model adopts a state space representation, and the trend component is estimated by a Kalman filter algorithm. The algorithm obtains one- or more-years-ahead prediction values using all past information from the data. Thus, depending on the number of years the investigator wants to consider as "the most recent", the expected trend for these years is estimated through the statistical procedure by using only information from the years prior to them. Forecast bands are estimated around the predicted trends for the recent years, and in the final step, an assessment is made on the extent to which observations from the most recent years fall outside these forecast bands. Those that do, may be identified as flagged observations. A procedure is also presented for assessing whether the combined information from all of the most recent observations form a pattern that deviates from the predicted trend and thus represents an unexpected tendency that may be flagged. In addition to form the basis for identifying time series that should be prioritized in an integrated ecosystem assessment, flagged observations can provide the basis for communicating with managers and stakeholders about recent ecosystem change. Applications of the framework are illustrated with two worked examples.

**Introduction**

Against a background of increasing impact from climate change and other anthropogenic drivers, causing elevated rates of change in marine ecosystems [1-6] leading to patterns of variability beyond the range of the Holocene [7-12], ecosystem-based management (EBM) is increasingly identified as a needed framework for management of marine socio-ecological systems [13]. Integrated ecosystem assessments (IEA) have been developed to provide the scientific basis for EBM [14], and numerous groups of scientists working with IEA have been established, such as the regional IEA groups within the International Council for the Exploration of the Sea (ICES, Walther and Möllmann (15)).

Among the core activities of IEA groups are analyses of time series to summarize changes that have occurred in recent decades in ecosystems, highlight possible connections between physical, biological, and human ecosystem components [14, 16]. Emphasis is put on keeping an open communication management and stakeholders [13, 17, 18]. As the groups typically have at their disposal a large number of time series [16, 19], it will often be necessary to prioritize a subset of them for more extensive analyses and communication purposes [20, 21]. Prioritization should preferably be done using a standardized framework applied to all time series. Here we present an approach which is based analyses of patterns of recent change, where the aim is to identify time series in which the most recent values deviate significantly from an expected trend, possibly indicating unexpected change. This should be of high relevance for IEA groups, as they are often challenged with understanding and interpreting recent change [14].

Our approach is based on first estimating trends of time series before assessing whether the most recent observations deviate significantly from these trends. Since temporal changes in ecosystems can take the form of long-term movements as well as short- or mid-term cyclic periods and noise components, different definitions of trends have been used in marine IEAs [16]. In the field of statistical time series analysis, the long-term movements are commonly classified as 'trends', while short- or mid-term cyclic periods are not, due to the different assumptions about the statistical properties. When investigating a trend in time series data, it can therefore be useful to separately identify non-stationary trends and stationary cyclic components. This decomposition is performed by a framework called 'structural time series modelling', which is using a state space representation where the state of each component is estimated by the Kalman filter algorithm [22]. The concept is basically different from applying an Autoregressive integrated moving average model to adjust with the aim of studying stationary processes from nonstationary time series data [23].

The Kalman filter algorithm can make one- or multistep-ahead predictions in the numerical procedure. The numerical procedure introduced in this paper uses prediction values and forecast uncertainty bands to assess the status of a recent observation, which determines whether the most recent observation follows the prediction or deviates from it [24], thus giving an indication whether change that is unexpected from the predicted trend, is occurring in the time series. We call the significant deviated observations a "Flagged Observation (FO)" and the approach "Flagged Observation analysis". The interpretation of a FO is not equivalent to the types of early warning signals that have been proposed with the aim of predicting critical transitions in marine populations or ecosystems [25, 26], nonlinear ecological change [27] or early warning signs based on theoretical framework in social-ecological networks [28], but is, as described above, a practical tool for IAE groups for prioritizing time series for in depth analyses, communication and other purposes.

In this article, we first introduce the numerical procedure and then demonstrate two examples using time series data for, respectively, the Atlantic Multi-decadal Oscillation (AMO) and the Norwegian Sea ecosystem.

**Statistical method**

The statistical method includes first a procedure for trend estimation and second a procedure for flagged observations analyses (FO analyses) based on multistep ahead prediction values. The output from these analyses can be used to identify observations (for example years) that deviate significantly from the expected trend. In addition, it can also be interesting to explore whether observations from the predicted years together form a pattern where all are consecutively either above or below the predicted trend in a way that is not expected. This is equivalent to asking whether there is an unexpected tendency for the most recent years. The probability for observing this is smaller than the probability of detecting an FO for a single observation. The details are given as below:

*Trend estimation procedure*

The observation model of a time series is given by

$$y(n) = t(n) + u(n), \quad n = 1, \cdots, N \quad , \tag{1}$$

where $t(n)$ is the trend component and $u(n)$ is the residual component at time step $n$, assuming Gaussian white noise. In this article, we introduce a stochastic trend model given by a $d$th-order difference equation model and a method for estimating the trend [29, 30].

The stochastic differential trend model is defined by the $d$th-order difference equation, which was posed as a smoothing problem by [31]. This model allows for more flexible trends than does the polynomial regression model. The stochastic trend model for $k$ variables is expressed in the following way:

$$\nabla^d t_i(n) = v_{t_i}(n), \quad i = 1, \cdots, k, \tag{2}$$

where $\nabla$ is a difference operator $\nabla t(n) = t(n) - t(n-1)$ and $v_{t_i}(n)$ is assumed to be a white noise sequence. If $d = 1$, $t(n) \approx t(n-1)$ and the trend is known as a random walk model. If $d = 2$, $t(n) - 2t(n-1) + t(n-2) \approx 0$ [29]. Provided that the variance of $v_{t_i}(n)$ is sufficiently small, $t_i(n)$ yields a smooth trend. We choose the second order difference stochastic model to estimate trend in this study.

The model can be represented in linear state space form [22, 29, 30], as

$$\begin{aligned} z(n) &= Fz(n-1) + Gv(n), \\ y(n) &= Hz(n) + w(n), \end{aligned} \tag{3}$$

where $z(n)$ is the state vector corresponding to $t(n)$, $v(n)$ is the system noise vector with mean 0 and unknown variance $\sigma_v^2$, $F$, $G$, and $H$ indicate integers or matrices, and $w(n)$ is observation error with mean 0 and unknown variance $\sigma_w^2$. The state $z(n)$ corresponds to the trend component, which we cannot directly observe from the data. The component is modelled by the $d$th-order differential equation model. Corresponding to the above $d$th-order difference equation, when $d = 1$ in equation (2),

$$z(n) = [t(n)], \quad F = G = H = 1.$$

When $d = 2$, the state vector and matrices are as follows:

$$z(n) = \begin{bmatrix} t(n) \\ t(n-1) \end{bmatrix}, \quad F = \begin{pmatrix} 2 & -1 \\ 1 & 0 \end{pmatrix}, \quad G = \begin{bmatrix} 1 \\ 0 \end{bmatrix}, \text{ and } H = \begin{bmatrix} 1 \\ 0 \end{bmatrix}.$$

The observation error corresponds to $u(n)$ in equation (1) in this case. The trend component is estimated with a Kalman filter, which is a powerful numerical algorithm that recursively operates the state estimation, prediction and filtering:

*prediction*:

$$\begin{aligned} z(n|n-1) &= Fz(n-1|n-1) \\ V(n|n-1) &= FV(n-1|n-1) + GQG' \end{aligned} \qquad (4)$$

*filtering*:

$$\begin{aligned} K(n) &= V(n|n-1)H'(HV(n|n-1)H' + R)^{-1} \\ z(n|n) &= z(n|n-1) + K(y(n) - Hz(n|n-1)) \\ V(n|n) &= (I - KH)V(n|n-1) \end{aligned} \qquad (5)$$

Here, $z(n|n-1)$ and $V(n|n-1)$ correspond to the conditional mean and conditional variance of the state, $R$ is set as the observation error, and $K$ is called the Kalman gain. Setting the initial state $z(1|0)$ and variance $V(1|0)$ as zero and the pre-determined system noise, the Kalman gain is calculated by the initial variance and observation noise, and the filtering value $z(1|1)$ is obtained by using observation $y(1)$ and the calculated gain from the filtering procedure (5). Then the next prediction values $z(2|1)$ and $V(2|1)$ are calcualted using $z(1|1)$ and $V(1|1)$ in the prediction procedure (4). The iterative calculation procedure for the state is continued until $n = N$. Recursive methods based on state space representations are known to be very efficient for calculating the likelihood functions of discrete-time Gaussian proceeses. The state space model and the Kalman filter provide an efficient method for the computation of the likelihood of the time series models [29]. In this case, the trend model includes the parameter vector $\theta = (d, \sigma_v^2, \sigma_w^2)$. The log-likelihood function $l(\theta)$ of the model is given by

$$l(\theta) = \sum_{n=1}^{N} \log f(y(n)|Y(n-1),\theta),$$

$$= \sum_{n=1}^{N} \log \left\{ \frac{1}{\sqrt{(2\pi)^2 \det \Sigma(n)}} \exp\left(-\frac{1}{2}\Delta y(n)'\Sigma(n)^{-1}\Delta y(n)\right) \right\},$$

where $Y(n-1) = (y(1), y(2), \cdots, y(n-1))$, $\Delta y(n) = y(n) - Hz(n|n-1)$, and $\Sigma(n) = H(n)V(n|n-1)H'(n) + \sigma_w^2$. The flexibility of the estimated trend depends on $\sigma_v^2$, which can be determined by maximum likelihood within an arbitry variance range. The variance $\sigma_w^2$ can be directly set to the variance of the observation. If it is necessary to compare differernt order differential stochastic model, the optimum differential order $d$ is identified by the AIC [32], which was formulated by the maximum log-likelihood and number of parameters for $d$ and the variances of system noise and observation noise, given by $\mathrm{AIC}(c) = -2l(\theta) + 2 \times \text{number of parameters}$.

After identifying the optimum trend model, the smoothed trend is estimated by a fixed-interval smoother algorithm [33]:

$$A(n) = V(n|n)F'V(n+1|n)^{-1}$$
$$z(n|N) = z(n|n) + A(n)(z(n+1|N) - z(n+1|n))$$
$$V(n|N) = V(n|n) + A(n)(V(n+1|N) - V(n+1|n))A(n)'$$

In this study, we set the differential order as $d = 2$ to obtain smooth trend for all time series data as introduced in Kitagawa and Gersch (33) and take a procedure finding an optimum $Q$ controlling variability for the trend estimation within a range.

*FO analysis by multistep-ahead prediction values*

Using the Kalman filter algorithm directly, it can give only one-ahead prediction. However, it can be expanded to multistep-ahead (*j*-ahead) prediction ($j > 1$). Let us consider a relevant situation. With the Kalman filter, one-ahead prediction for $z(n+1)$ is obtained by $z(n+1|n)$ and variance $V(n+1|n)$. If the data $y(n+1)$ are not observed, the calculation is formally conducted by assuming

$Y(n+1) = Y(n)$, where $Y(n) = (y(1), y(2), \cdots, y(n))$. Accordingly, it is clear that $z(n+1|n+1) = z(n+1|n)$ and $V(n+1|n+1) = V(n+1|n)$. Then, two-ahead prediction $z(n+2|n)$ and variance $V(n+2|n)$ are obtained by $z(n+1|n)$ and $V(n+1|n)$. In general, we assume $Y(n) = Y(n+1) = \cdots = Y(n+j)$ to obtain the $j$-ahead prediction, and the prediction step is iteratively conducted $j$ times. The algorithm used to predict states $z(n+1), z(n+2), \cdots, z(n+j)$, based on the data $Y(n)$ observed until time point $n$, is expressed as follows:

$$z(n+i|n) = Fz(n+i-1|n),$$
$$V(n+i|n) = FV(n+i-1|n)F' + GQG', \quad i = 1, \cdots, j.$$

Now, let the mean and variance for the prediction $y(n+j)$ of the data denote $\mathrm{E}(y(n+j)|Y(n))$ and $\mathrm{Cov}(y(n+j)|Y(n))$, where $\mathrm{E}(\cdot)$ and $\mathrm{Cov}(\cdot)$ are notations for expectation and variance-covariance matrix (but in this study only variance because the data are univariate). Using the observation equation (3), the mean of $y(n+j)$ is expressed by

$$y(n+j|n) = \mathrm{E}(Hz(n+j) + w(n+j)|Y(n)) = Hz(n+j|n). \tag{6}$$

The variance of $y(n+j)$ is given by

$$\begin{aligned}
d(n+j|n) &= \mathrm{Cov}(Hz(n+j) + w(n+j)|Y(n)), \\
&= H\,\mathrm{Cov}(z(n+j)|Y(n))H' + H\,\mathrm{Cov}(z(n+j), w(n+j)|Y(n)) \\
&\quad + \mathrm{Cov}(w(n+j), z(n+j)|Y(n))H' + \mathrm{Cov}(w(n+j)|Y(n)), \\
&= HV(n+j|n)H' + R(n+j).
\end{aligned} \tag{7}$$

Therefore, the prediction distribution for $y(n+j)$ based on data $Y(n)$ is a normal distribution with mean $y(n+j|n)$ and variance $d(n+j|n)$ or a standard deviation $\sqrt{d(n+j|n)}$. The forecast bands (FBs), e.g. mean ± 1.96 (~2) × standard deviation that corresponds to around 95-96% forecast interval [34], are easily calculated by equations (6) and (7). Note that we define the values given by multistep-ahead prediction procedure as 'forecast value' because it is calculated by using the previous data in the algorithm.

*Assessing unexpected tendencies - joint probability for detected FO*

As the time series data is sampled as one sample at one time point, the joint probability that all of the most recent years fall either above or below the predicted trend (i.e., whether there is an unexpected tendency for the most recent years) should be calculated by random variables, which is generated from a normal distribution. We provide the following procedure to calculate the joint probabilities in this way:

1. Generate 10000 random number set $r_j$ by a normal distribution with mean $y(n+j\,|\,n)$ and variance $d(n+j\,|\,n)$ of the prediction values at $n+j$;

2. The probability value $p_j$ is calculated by the formula

$$p_j = \begin{cases} \dfrac{\#\{r_j \leq y(n+j\,|\,n)\}}{10000} & \text{if } y(n+j\,|\,n) \text{ is upper over FB} \\ \dfrac{\#\{r_j \geq y(n+j\,|\,n)\}}{10000} & \text{if } y(n+j\,|\,n) \text{ is lower under FB} \end{cases} ;$$

3. The joint probability values $P(p_1,\cdots,p_J)$ are calculated for $j=1,\cdots,J$ by

$$P(p_1,\cdots,p_J) = p_1 \times \cdots \times p_J; \text{ and}$$

4. The procedure from 1 to 3 is iterated for 1000 and the averaged joint probability values are calculated by

$$\overline{P}(p_1,\cdots,p_J) = \frac{1}{1000}\sum\nolimits_{b=1}^{1000} P_b(p_1,\cdots,p_J).$$

A conceptual outline of this study's analysis procedure for FO analysis is given in Fig 1. The numerical procedure is implemented using MATLAB code [35], which is summarized in Supplementary folder.

**Illustrative examples**

The dataset for the Atlantic Multi-decadal Oscillation (AMO) is based on index monthly raw data [36], while for the Norwegian Sea ecosystem, we use the yearly data assembled by the ICES integrated ecosystem assessment working group for the Norwegian Sea (WGINOR, ICES (37)). Abbreviations for the Norwegian Sea data used in this article is summarized in Supplementary Table1. In the examples, we do not attempt to fully interpret any FOs revealed, but comment on the possible background for some of them to illustrate the context for further work in IEA groups.

*The Atlantic Multi-Decadal Oscillation (AMO)*

AMO is a pronounced signal of climate variability in the North Atlantic Sea surface temperature (SST) [38]. The monthly data recorded from December in 1869 to March in 2021 has been published under the NCAR CLIMATE data guide [39]. We extracted monthly raw data for the period 1980-2020, from which we calculated annual means, giving a time series with 31-time points.

To illustrate how inferences may differ for different time periods, both seven-years and three-years predictions are shown for this example. Thus, we first set the specific time point $j$ as 2014 and 2017, respectively. The stochastic difference trend model was applied to the data until $j$-1 time points (that is, 2013 and 2016). To optimize $Q$ for the model, we set $0.05 \leq Q \leq 0.5$ as a search range. The calculated maximum log-likelihood and optimum $Q$ for each dataset are summarized in Table 1a. The details for the log-likelihood in the range of Q are summarized in Table 2.

Using the parameters of the model, the Kalman filter algorithm was run to calculate seven-years- and three-years-ahead predictions. Fig 2 presents the outputs of this.

For the case of seven-years-ahead prediction, none of the observations for the most recent years fall outside the 95% or 80% FBs, but the observation for 2015 fall outside the 70% FB. Thus, only a single possible FO is identified when looking at each of the seven most recent years individually, and this is due to a marked decrease in SST in 2015 that contrasts the slightly increasing trend predicted for 2014-2020. However, all observations for the seven most recent years fall below the predicted trend (Fig 2). The joint probability for this pattern is small (Table 2), suggesting that there is a tendency in the data that differs from the predicted trend and thus represents a FO. For the case of three-years-ahead predictions, none of the observations from the most recent years fall outside any of the FBs, and they are spread evenly around the predicted trend (Fig 2). Thus, while there are ample indications that the seven most recent observations deviate from the trend predicted for the last seven years, no such pattern is seen for the trend predicted for the last three years, suggesting that an assessment group may need to look differently at change over these two time periods.

*Norwegian Sea ecosystem*

The Norwegian Sea is located West and Northwest of Norway, bordered by the North Sea and the Atlantic Ocean to the south, the Greenland Sea to the west and the Arctic Ocean and Barents Sea to the north and east. It is a deep-sea area with three species of mainly planktivorous pelagic fish making up the economically most important fish stocks: mackerel (*Scomber scombrus*), Norwegian spring-spawning herring (*Clupea harengus*) and blue whiting (*Micromesistius poutassou*). Ocean currents are dominated by relatively warm and saline Atlantic water masses flowing in from the south and colder and fresher Arctic water masses flowing in from the northwest [40]. There is considerable negative density dependence acting on biomass within the three pelagic fish stocks, presumably through intraspecific competition over food [41]. While there are also indications of competition among the stocks, most strongly between mackerel and herring [41], other work has suggested that interspecific competition is less significant [42]. The combined biomass of the three species has increased over the last decades while zooplankton biomass has declined, and it has been hypothesized that the pelagic fish biomass may have exceeded the carrying capacity of the system [21]. The climate has historically varied between cold and warm phases, with plankton and fish productivity tending to increase in the warmer phases [43]. Here, we analyse time series on spawning stock biomass, recruitment and growth (age and weight at age 6) for the three pelagic fish stocks, zooplankton biomass, and three key variables for the physical environment: heat content, freshwater content and the North Atlantic Oscillation index. The observations have been recorded annually, although the starting/ending years of observation vary among the time series.

For the current work, we used time series with different start years and 2019 as the last year [37]. As one of the main aims of IEAs in the Norwegian Sea has been to provide background information for advisory work for operational fisheries management [44, 45], change over a short period of the most recent years is typically of interest. The conditions for making the prediction were therefore set to three-years-ahead predictions for 2017-2019 using the data observed up to 2016. The calculated maximum likelihood, and optimum $Q$ for each data are summarized in Table 1. The $\sigma_w^2$ was calculated using the observations until 2016.

Using the parameters of the model, the Kalman filter algorithm was run to calculate three-years-ahead predictions. Fig 3 presents the outputs of this. Looking at variables for the physical environment, FOs for individual years were observed for relative freshwater content (RFW), where the observation for 2018 fall outside the 80% FB and for 2019 outside the 95% FB. In addition, observations for all of the last three years fall well above the predicted trend, and the joint probability for this pattern is low (Table 2), indicating an unexpected tendency in the data when compared with the expected trend. While freshening of the Norwegian Sea had been going on for nearly a decade before 2019 [46], these unexpected increases in freshwater content point to a recent intensification of this that might require the attention in IEAs of the Norwegian Sea.

For zooplankton biomass (ZooB), two of the most recent years fall above the 80% FB and above the 70% FB, with a low overall probability of the pattern (Table 2), indicating, again, an unexpected upward tendency in the data. This suggests that an IEA should pay attention to a possible recovery of the zooplankton biomass, which declined sharply in the early 2000s and remained at low levels in the following years [18].

Looking at variables for pelagic fish stocks, little evidence for unexpected changes is seen for herring and mackerel. For the four variables related to mackerel, no years fall outside any of the FBs and observations generally lie relatively close to or on both sides of the predicted trends (Fig 3). A similar pattern is seen for herring, except one estimate of recruitment falling above the 70% FB (Fig 3). As pelagic fish recruitment is highly variable, a single observation falling outside the expected trend may not be reason for flagging this variable for prioritization in an IEA.

A different picture emerges for blue whiting. For spawning stock biomass (BWB), two of the three most recent years fall above the 70% FB and the third year also well above the predicted trend (Fig 3). The joint probability of this pattern is low (Table 2), suggesting that there is a tendency for an increase in biomass beyond the expected. At the same time, there are indications of unexpected declines in blue whiting individual growth, shown for weight at age 6 (BWW), where all observations fall below the 70% FB (Fig 3) and the joint probability for the pattern is low (Table 2). These changes may be linked, as increases in biomass tend to be associated with decreases in growth, possibly though intraspecific competition over food [41]. In addition, for blue withing recruitment (BWR), two observations fall below the 70% FB and one below the 80% FB (Fig 3) with a low joint probability for the overall pattern (Table 2), indicating that the decline in recruitment for the most recent years represent an unexpected tendency. Although pelagic fish recruitment remains hard to forecast (e.g. [47]), it is interesting to note that considerable progress has been made in predicting variation in the geographical location of blue whiting spawning habitat, which may be linked to recruitment success [48-50], thus offering a possible avenue for more detailed assessments and studies following up the changes in blue whiting recruitment.

**Discussion**

We have introduced a time series analysis procedure for making predictions for a specific time period using a structural time series model including a trend model. Based on this, we have outlined a framework for investigating whether the most recent observations deviate from the predicted trend for this time period and thus represent possible flagged observations (FOs). This includes assessing both whether single years represent FOs or whether all of the observations from the recent years together represent an unexpected tendency that is classified as a FO. The trend was estimated using a stochastic trend model and observed time series data, and the specific-years-ahead predictions were systematically calculated according to the iterative procedure of the Kalman filter algorithm. The statistical analysis is followed by a qualitative evaluation of each FO, where it may be decided to follow some of them up by more detailed analyses within an integrated ecosystem assessment (IEA). We note that applications may also extend beyond IEAs to other areas of science and advisory processes where an overview of recent change is required across multiple time series.

The time series available from marine ecosystems that are relevant for analyses described here are typically short (i.e., < 50 time points) [19, 51]. This puts constraints on the types of time series analyses that can be performed. For example, null hypothesis testing using a frequentist statistical approach can produce misleading results, including false positive and negative results [52]. The procedure described here does not include null hypothesis statistical testing, and the type of structural time series model used by us is not based on a frequentist framework but corresponds to a Bayesian approach [53], which may properly assess trends in short time series that cannot be analysed using a frequentist approach [52]. An alternative method to the one used here could have been the Box Jenkins model, which transforms a non-stationary mean time series to a stationary process [54]. However, effective fitting using this method, again, requires longer time series than what is normally available in marine ecosystems [19, 51]. Thus, for short time series, the Bayesian framework used here appears to be more robust than alternative approaches.

To study and assess recent change, IEA groups often rely on examinations of anomaly plots. In such plots, recent change appears as deviations from a long-term mean, often estimated for the whole length of the time series (see e.g. Bulgin, Merchant (55) for an application to global sea surface temperatures, SST). By focusing on deviations from the expected trend for the most recent years (where the expected trend is estimated by using information from the whole time series), FO analysis can provide a different perspective of recent change. For example, using a seven-years prediction, the FO analysis indicates that there is a tendency in North Atlantic Sea SST towards a more negative trend than what should be expected for the last 7 years of the time series. We argue that the same interpretation is less evident from an anomaly plot of the same time series (see Supplementary Fig 1). Thus, while positive and negative values indicate how observations deviate from a *constant* mean value in an anomaly plot, the trend changes through time, making it harder to assess how the most recent observations deviate from the trend that should be expected for these recent years. Recognizing that anomaly plots are important for a large range of purposes within IEAs, we emphasize that FO analysis can provide useful additional information for the practical work in IEA groups, in particular in the light of the challenge they are often faced with of understanding and assessing the most recent development of an ecosystem [14].

Since the cumulative output of FO analysis also aim at giving a sweeping overview of the recent dynamics of all the measured elements in an ecosystem by highlighting the variables that exhibit unexpected change while at the same time showing trends and data for those that do not, the approach should be useful for facilitating the necessary dialogue between scientists and stakeholders about recent ecosystem change within the process of an IEA. Such overviews can also contribute to the scientific output used to educate and inform the public and the political system during parts of policymaking processes related to for example ecosystem-based management [56].

# Acknowledgements

We would like to thank Benjamin Planque, who motivated us to consider this approach to analysing the time series data compiled by the ICES integrated ecosystem assessment working groups and Mette Mauritzen and Daniel Howell for comments on an earlier draft of the paper. This study was carried out as part of the project "Sustainable multi-species harvest from the Norwegian Sea and adjacent ecosystems", funded by The Research Council of Norway (pr. nr. 299554).

Table 1. Calculated log-likelihood (LL) and the optimum $Q$ by applying stochastic trend model for $d = 2$ for the AMO and Norwegian Sea ecosystem datasets.

| Dataset | Variable | Q | Maximum log-likelihood |
|---|---|---|---|
| AMO | AMOS | 0.50 | -51.2 |
| Norwegian Sea ecosystem (WGINOR) | RHC | 0.05 | -82.8 |
| | RFW | 0.08 | -82.6 |
| | NAO | 0.05 | -161.3 |
| | ZooB | 0.09 | -29.1 |
| | MacB | 0.12 | -43.3 |
| | MacR | 0.06 | -42.1 |
| | MacW | 0.07 | -44.1 |
| | MacL | 0.07 | -71.8 |
| | HerB | 0.06 | -125.7 |
| | HerR | 0.05 | -42.6 |
| | HerW | 0.11 | -87.1 |
| | HerL | 0.08 | -93.7 |
| | BWB | 0.13 | -42.4 |
| | BWR | 0.16 | -48.7 |
| | BWW | 0.09 | -43.8 |
| | BWL | 0.08 | -57.3 |

Table 2. Averaged p-values and joint probability for the prediction and observation for AMOS, RFW, ZooB, BWB, BWR, and BWW.

| Year | AMOS | RFW | ZooB | BWB | BWR | BWW |
|---|---|---|---|---|---|---|
| 2014 | 0.19 | | | | | |
| 2015 | 0.14 | | | | | |
| 2016 | 0.29 | | | | | |
| 2017 | 0.42 | 0.22 | 0.12 | 0.13 | 0.1003 | 0.12 |
| 2018 | 0.22 | 0.061 | 0.28 | 0.16 | 0.1237 | 0.11 |
| 2019 | 0.28 | 0.023 | 0.13 | 0.29 | 0.1415 | 0.13 |
| 2020 | 0.42 | | | | | |
| Joint probability | 8.24e-05 | 3.06e-04 | 0.0045 | 0.0063 | 0.0018 | 0.0016 |

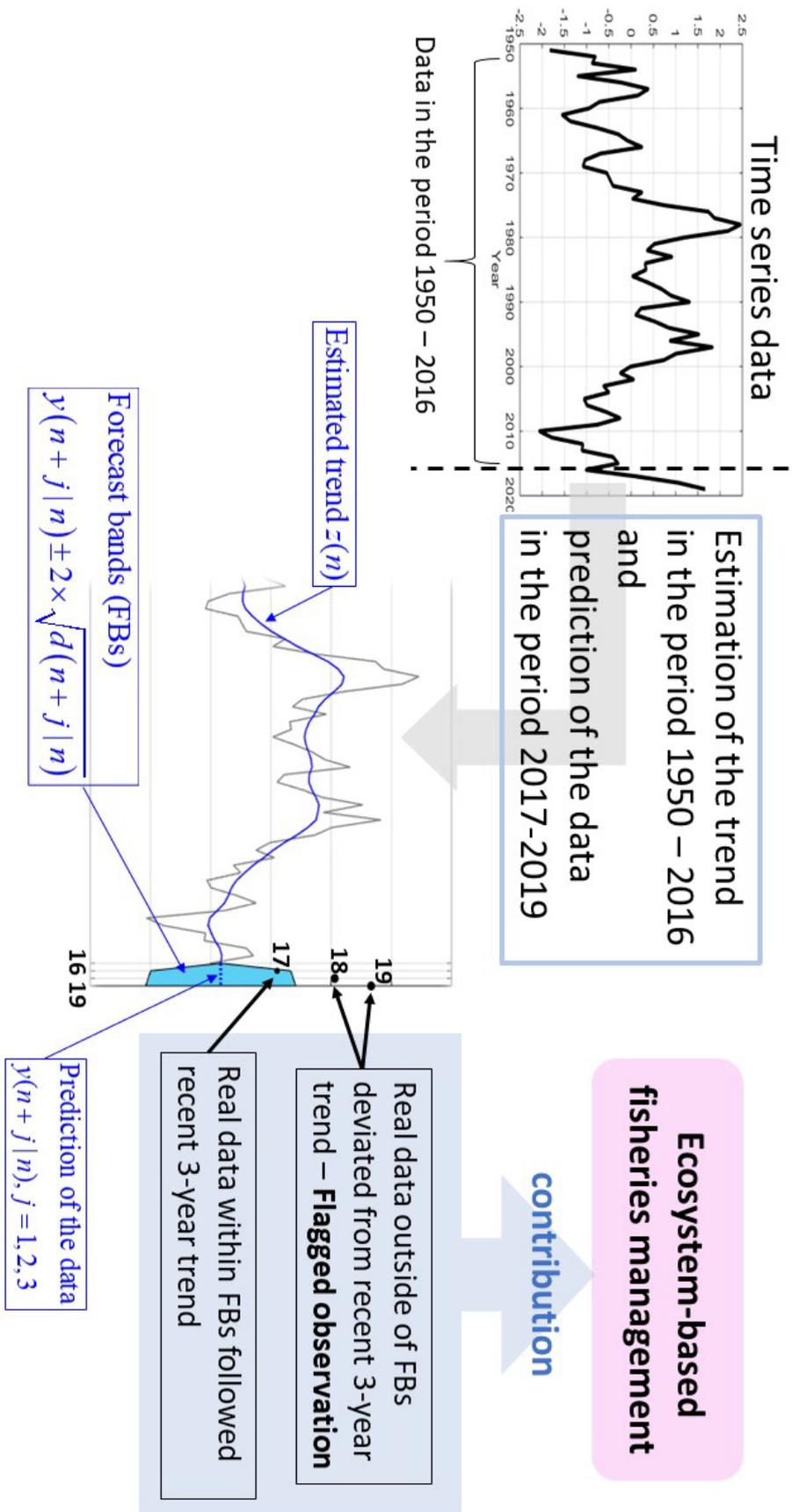

Fig1. Outline of proposed Flagged observation (FO) analysis

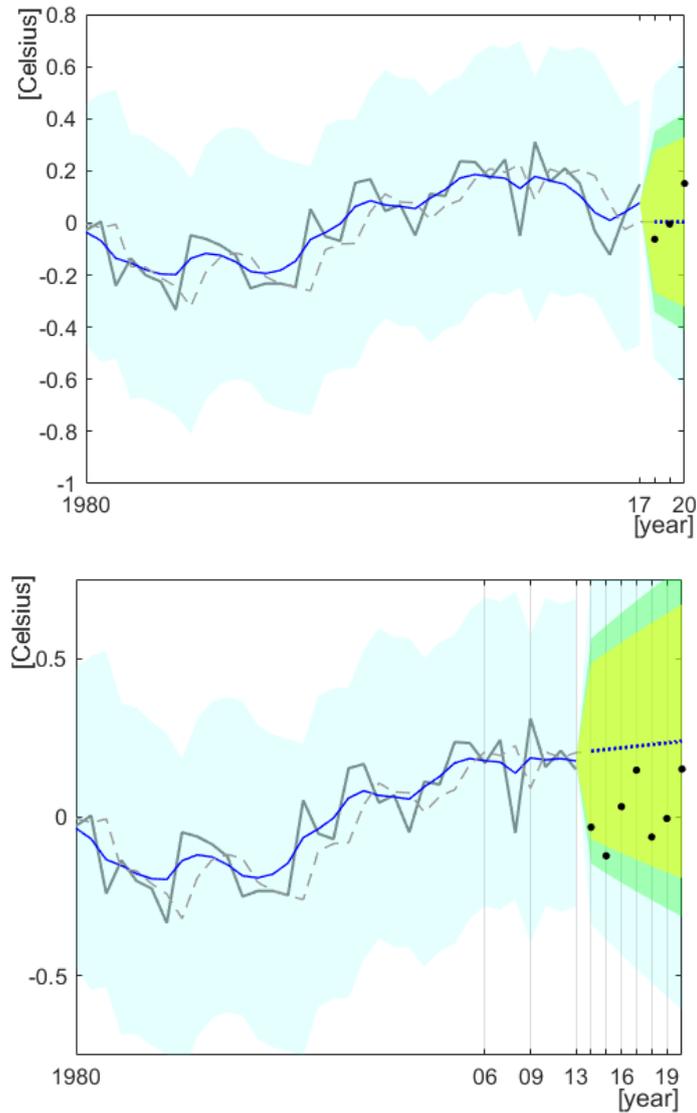

Fig 2. The estimated three-years (upper) and seven-years (lower) prediction values of sea surface temperature and the most recent observations for the dataset on the Atlantic Multi-decadal Oscillation. The solid grey lines indicate the observations used for making the forecast values and the black points indicate the observations that were plotted for comparison with the prediction values (dotted blue line). The dotted grey line presents the prediction value $Hz(n|n-1)$ and the solid blue lines present the smoothed trend estimates obtained by a fixed-interval smoother algorithm. The band coloured by light-blue presents the 95% FB, 80% and approximately 70% FBs, which upper and lower limits were calculated by mean $y(n+j|n)$ ± 1.96 (1.28, 1) × standard deviation $\sqrt{d(n+j|n)}$, where $j=1,2,\cdots,7$ or $j=1,2,3$. The observations in the years applying the prediction are shown with black points.

Climate:

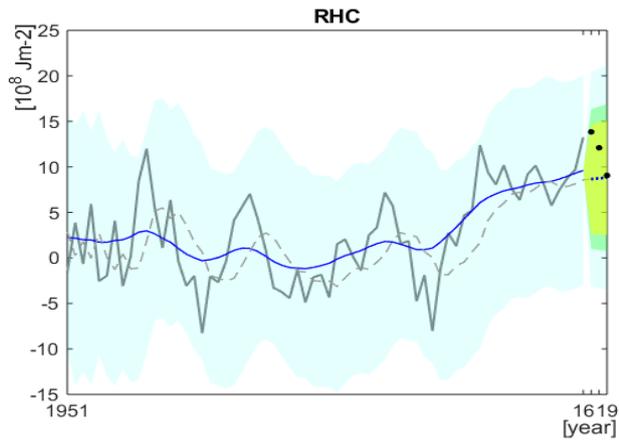

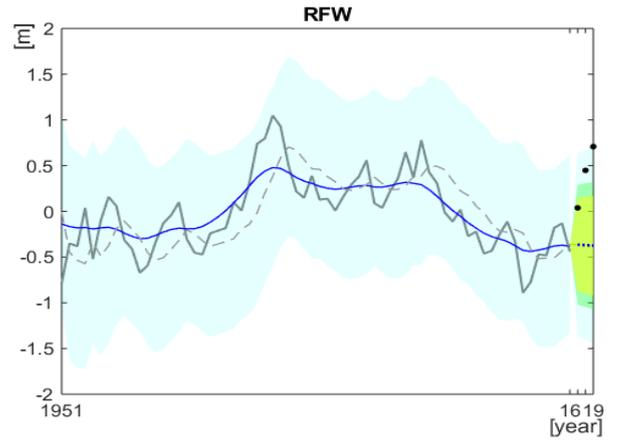

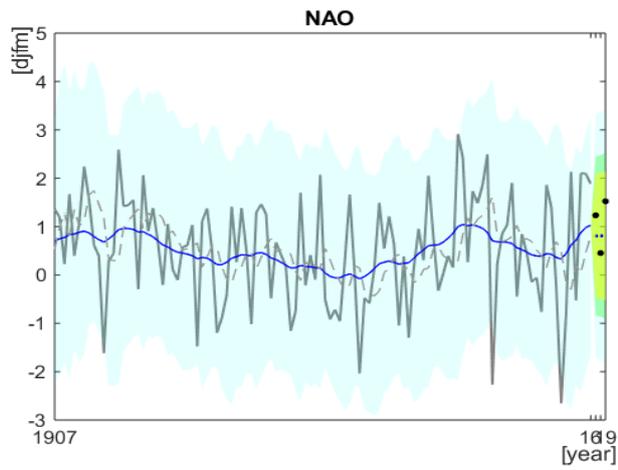

Plankton:

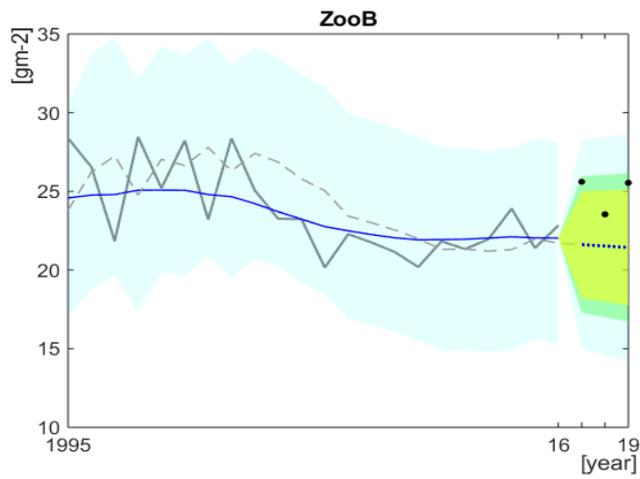

Fig 3 *Continued.*

Pelagic fish:

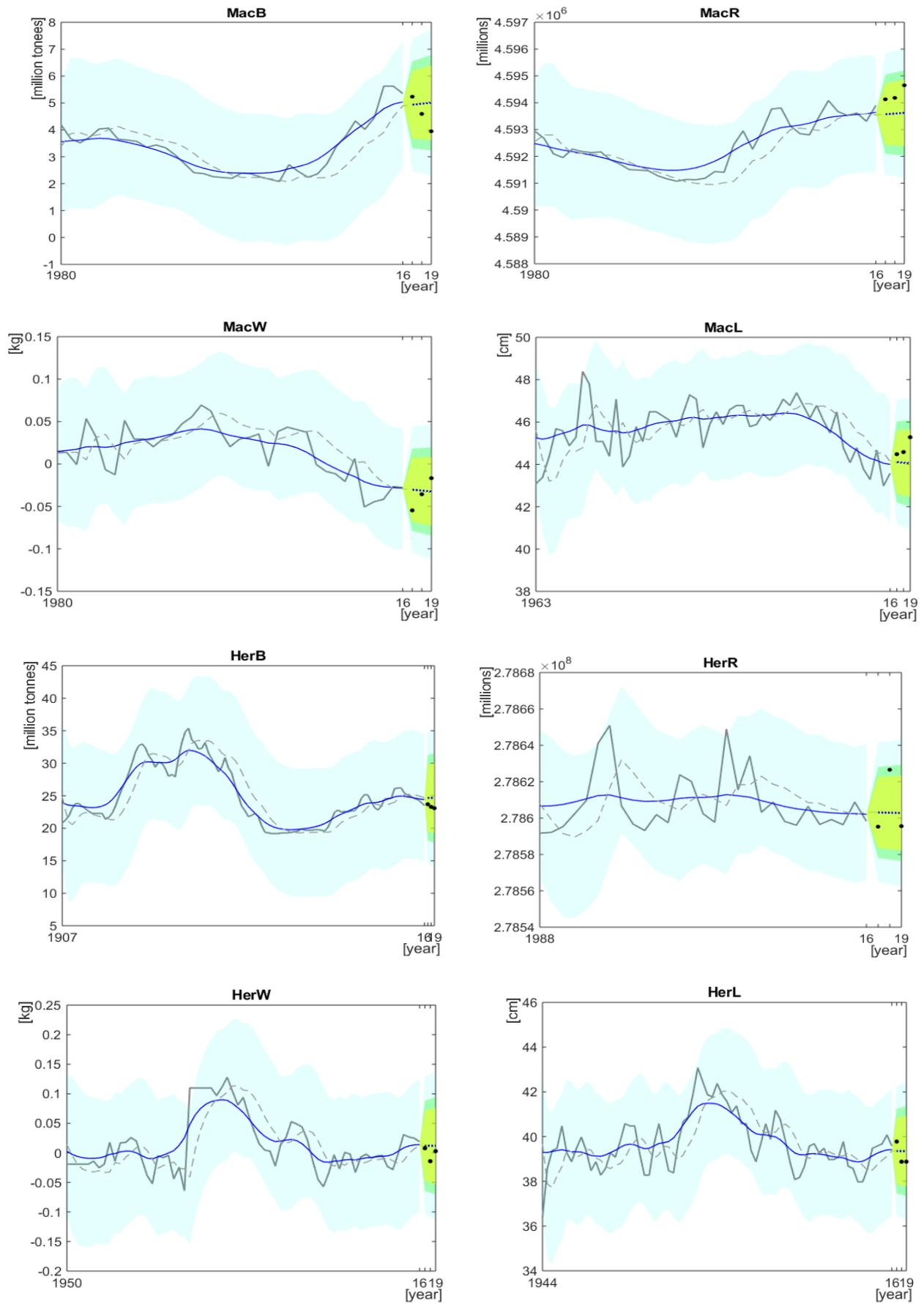

Fig 3 *Continued*.

Pelagic fish:

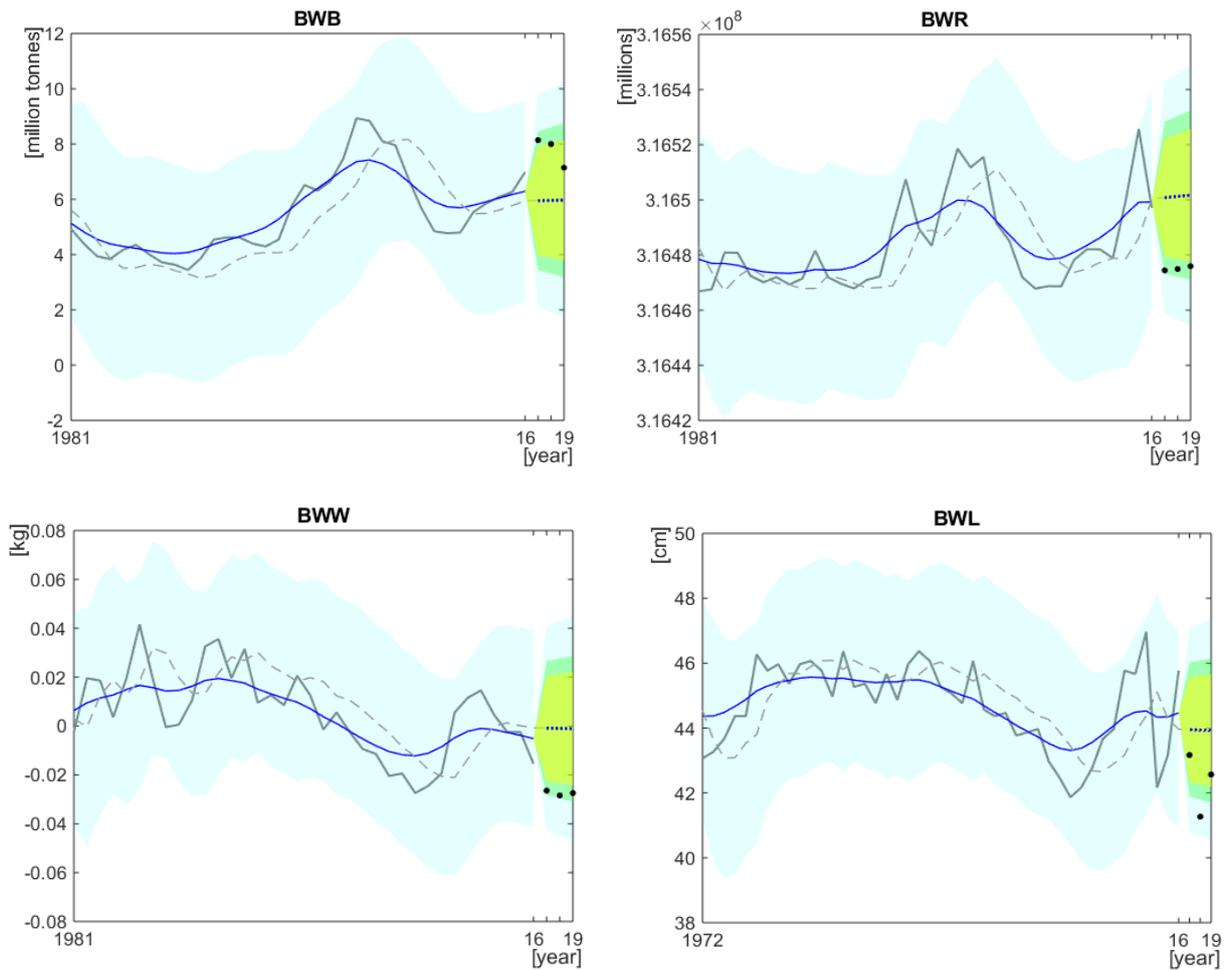

Fig 3. The estimated three-years prediction values and the three most recent observations for data on variables related to climate, plankton and pelagic fish in the Norwegian Sea ecosystem. The solid grey lines indicate the observations used for making the forecast values and the black points indicate the observations that were plotted for comparison with the prediction values (dotted blue line). The dotted grey line presents the prediction value $Hz(n|n-1)$ and the solid blue lines present the smoothed trend estimates obtained by a fixed-interval smoother algorithm. The band coloured by light-blue presents the 95% FB, 80% and approximately 70% FBs, which upper and lower limits were calculated by mean $y(n+j|n)$ ± 1.96 (1.28, 1) × standard deviation $\sqrt{d(n+j|n)}$, where $j=1,2,3$. The observations in the years applying the prediction are shown with black points.

Supplementary Fig 1

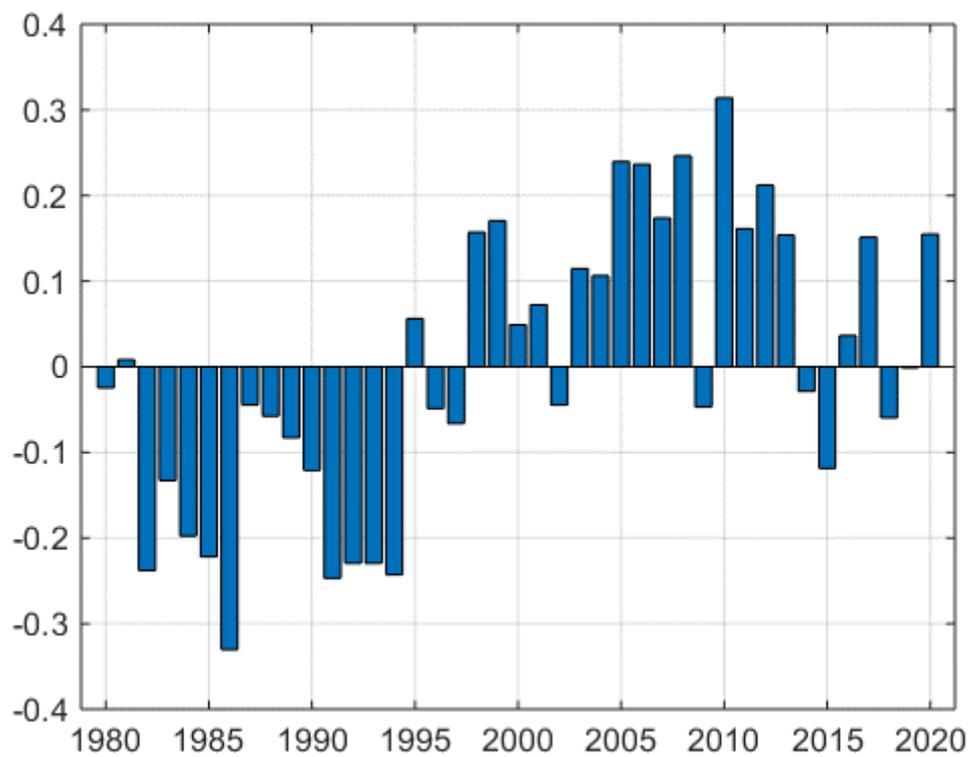

Supplementary Fig 1. Bar plots for anomalies of AMOS's yearly data from 1980 to 2020. The y-axis indicates the value subtracting mean value of yearly data from 1980 to 2020. The x-axis indicates year. The negative/positive values correspond lower/higher temperature to the mean value.

Supplementary Table. Abbreviations used in figures and tables

| Type | Abbreviation in figures and tables | Explanation of relevant data |
|---|---|---|
| Climate | RHC | Relative heat content in $10^8$ Jm$^{-2}$ |
|  | RFW | Relative freshwater content in m |
|  | NAO | North Atlantic Oscillation expressed as djfm |
| Zooplankton | ZooB | Total zooplankton biomass the Norwegian Sea in in May, g m-2 |
| Pelagic fish | MacB | Spawning stock biomass of Mackerel in million tonnes |
|  | MacR | Recruitment of Mackerel per year class at age 0 in millions |
|  | MacW | Weight of Mackerel at age 6 in the stock (in kg) |
|  | MacL | Length of Mackerel at age 6 in cm |
|  | HerB | Spawning stock biomass of Herring in million tonnes |
|  | HerR | Recruitment of Herring per year class at age 2 in millions |
|  | HerW | Weight of Herring at age 6 in the stock (in kg) |
|  | HerL | Length of Herring at age 6 in cm |
|  | BWB | Spawning stock biomass of Blue whiting in million tonnes |
|  | BWR | Recruitment of Blue whiting per year class at age 1 in millions |
|  | BWW | Weight of Blue whiting age 6 in the catch (in kg) |
|  | BWL | Length of Blue whiting at age 6 in cm |